# Terahertz spin-to-charge current conversion in stacks of ferromagnets and the transition-metal dichalcogenide NbSe$_2$

Lukáš Nádvorník[1,2,*], Oliver Gueckstock[1,3], Lukas Braun[3], Chengwang Niu[4,5], Joachim Gräfe[6], Gunther Richter[6], Gisela Schütz[6], Hidenori Takagi[7], Tom S. Seifert[1], Peter Kubaščík[2], Avanindra K. Pandeya[8], Abdelmadjid. Anane[9], Heejun Yang[10], Amilcar Bedoya-Pinto[8], Stuart S.P. Parkin[8], Martin Wolf[3], Yuriy Mokrousov[4], Hiroyuki Nakamura[6,11], Tobias Kampfrath[1,3,*]

1. Department of Physics, Freie Universität Berlin, 14195 Berlin, Germany
2. Faculty of Mathematics and Physics, Charles University, 121 16 Prague, Czech Republic
3. Department of Physical Chemistry, Fritz Haber Institute of the Max Planck Society, 14195 Berlin, Germany
4. Peter Grünberg Institut, Forschungszentrum Juelich, D-52425 Juelich, Germany
5. School of Physics, State Key Laboratory of Crystal Materials, Shandong University, 250100 Jinan, China
6. Max Planck Institute for Intelligent Systems, 70569 Stuttgart, Germany
7. Max Planck Institute for Solid State Research, 70569 Stuttgart, Germany
8. Max Planck Institute for Microstructure Physics, 06120 Halle (Saale), Germany
9. Unité Mixte de Physique, CNRS, Thales, Univ. Paris-Sud, Université Paris-Saclay, 91767 Palaiseau, France
10. Department of Physics, Korea Advanced Institute of Science and Technology (KAIST), Daejeon 34141, Korea
11. Physics Department, University of Arkansas, Fayetteville, Arkansas 72701, USA

* Email: tobias.kampfrat@fu-berlin.de, nadvornik@karlov.mff.cuni.cz

**Abstract**

Transition-metal dichalcogenides (TMDCs) are an aspiring class of materials with unique electronic and optical properties and potential applications in spin-based electronics. Here, we use terahertz emission spectroscopy to study spin-to-charge current conversion (S2C) in the TMDC NbSe$_2$ in ultra-high-vacuum-grown F|NbSe$_2$ thin-film stacks, where F is a layer of ferromagnetic Fe or Ni. Ultrafast laser excitation triggers an ultrafast spin current that is converted into an in-plane charge current and, thus, a measurable THz electromagnetic pulse. The THz signal amplitude as a function of the NbSe$_2$ thickness shows that the measured signals are fully consistent with an ultrafast optically driven injection of an in-plane-polarized spin current into NbSe$_2$. Modeling of the spin-current dynamics reveals that a sizable fraction of the total S2C originates from the bulk of NbSe$_2$ with the same, negative, sign as the spin Hall angle of pure Nb. By quantitative comparison of the emitted THz radiation from F|NbSe$_2$ to F|Pt reference samples and the results of *ab-initio* calculations, we estimate that the spin Hall angle of NbSe$_2$ for an in-plane polarized spin current lies between -0.2% and -1.1%, while the THz spin-current relaxation length is of the order of a few nanometers.

## 1. Introduction

Transition-metal dichalcogenides [1] (TMDCs) are an emerging class of materials with a C-TM-C stacking structure, where C and TM, respectively, denote a chalcogen atom (such as Se or S) and a transition-metal atom (such as Nb, W or Mo). In the last decade, TMDC monolayers (MLs) have attracted considerable attention [2-12] owing to their unique combination of electronic and optical properties. The hexagonal crystal structure of such quasi-two-dimensional materials implies inequivalent K-valleys in their electronic band structure, which give rise to the valley degree of freedom and valley-based electronic functionalities (valleytronics) [13]. The TM atoms provide large spin-orbit coupling (SOC) [14], which leads to further unique properties such as spin-valley locking [15], selective excitation of valley and spin polarizations [16], large spin-orbit torque [17] and a large valley Hall effect [18]. The combination of valley-dependent physics and pronounced SOC makes TMDCs excellent candidates for spintronic, opto-spintronic and valleytronic applications.

Most spin-related features of TMDCs, in particular optical spin injection [16], are uniquely associated with the out-of-plane spin orientation. In contrast, in-plane spin dynamics have received attention only in recent works [17,19,20], in which spin pumping was used to measure spin-orbit torque or the interface-related Rashba-Edelstein effect in ferromagnet|TMDC bilayers. On the other hand, only a few studies directly addressed the spin Hall effect [21], the conversion of a longitudinal charge current into a perpendicular spin current, or its inverse [22]. The in-plane spin geometry is particularly favorable for the observation of the SHE in TMDCs as the valley Hall effect is not operative in this geometry.

Terahertz (THz) emission spectroscopy is an excellent tool to study such spin transport phenomena and spin-to-charge current conversion (S2C) on their natural, i.e., ultrafast, time scales in, for example, fully metallic [23-28] or insulating-magnet|normal-metal heterostructures [29-31] and newly emerging materials like TMDCs [22,32]. An interesting application is the versatile optical generation of broadband THz electromagnetic pulses [24,33,34]. As shown in Figure 1a, the operation of such metallic spintronic THz emitters is based on the optically triggered generation of a spin voltage (spin accumulation) in the magnetic layer [35,36]. It drives a spin current that is transformed into a transverse charge current by ultrafast S2C, resulting in the emission of an electromagnetic pulse with frequencies extending into the THz range [24,36-38]. The THz emission approach (Figure 1a) is also useful to approximately determine the relative strength of ultrafast S2C of materials in a contact-free and rapid manner [24,39-41].

In this work, we address ultrafast spin transport and S2C in F|TMDC stacks, where F is a metallic ferromagnetic layer F of Fe or Ni, and the TMDC layer is $NbSe_2$. Following optical excitation of F|$NbSe_2$, we observe a sizable emission of broadband THz pulses. The temporal dynamics and $NbSe_2$-thickness dependence of the emitted THz electric field are fully consistent with the notion that optical excitation drives an ultrafast in-plane-polarized spin current into the bulk of $NbSe_2$. A qualitative comparison to a spin-transport model indicates that the TMDC, the ferromagnet and, possibly, their interface contribute significantly to the total S2C. A comparison to F|Pt reference samples allows us to make a quantitative estimate of the spin Hall angle and spin current relaxation length of $NbSe_2$ at THz frequencies, yielding –(0.2-1.1)% and 0.9-6 nm, respectively. *Ab-initio* calculations are fully consistent with the obtained range of spin Hall angles.

## 2. Results and discussion

*2.1 Samples*

The basic structure F|N of our samples is shown in Figure 1a, where F and N is a ferromagnetic and normal-metal layer, respectively. For clarity, all samples are always shown in this form even though the substrate is on the right-hand side. Atomically thin films of N=$NbSe_2$ are grown by hybrid pulsed-laser deposition (hybrid-PLD) [42,43] on double-side-polished sapphire substrates. Images of reflective high-energy electron diffraction (RHEED) taken after the $NbSe_2$ growth (Figure 1b) indicate two-dimensional layer-by-layer growth. The TMDC layer is covered by a layer of a ferromagnetic metal F (thickness of 3 nm), and finally capped by an AlOx protection layer (2 nm). Details on the sample preparation can be found in the Methods Section.

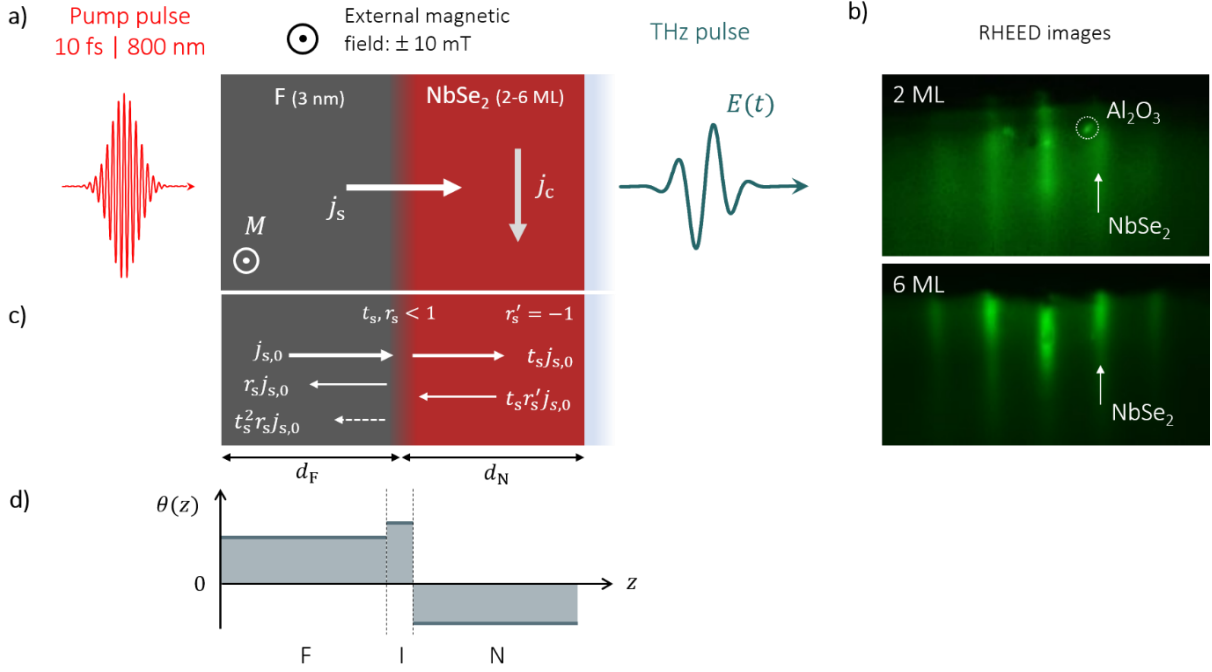

**Figure 1.** Schematic of laser-induced THz emission from F|NbSe₂ stacks. a) A F|NbSe₂ thin-film stack consisting of a metallic ferromagnetic layer F=Fe or Ni with magnetization **M** on top of a layer of the metallic TMDC NbSe₂ (thickness of 2-6 ML) is excited by a femtosecond laser pulse. The resulting spin voltage injects an ultrashort spin-polarized spin current $j_s$ from F into the adjacent NbSe₂. Spin-orbit coupling converts $j_s$ into a perpendicular charge current $j_c$. The resulting THz electric field $E(t)$ is detected by electro-optic sampling yielding a THz signal $S(t)$. b) RHEED images of 2 ML and 6 ML of NbSe₂, taken *in-situ* during the hybrid-PLD growth. c) Schematic of the spin-current echoes flowing through the F|N interface with interface transmission and reflection coefficients $t_s, r_s < 1$ and $r_s' = -1$ at the interface to the substrate. d) Illustration of regions with notable S2C, each characterized by its own spin Hall angle $\theta(z)$.

Note that S2C can, in principle, take place in any plane located at a depth $z$ of the F|NbSe₂ stack (Figure 1d). To extract the charge-current amplitude inside the NbSe₂ layer, we study samples with F=Fe or Ni, which feature opposite spin Hall angles ($\theta_{Fe} < 0$ and $\theta_{Ni} > 0$) [44], and NbSe₂ thicknesses $d_N$ between 2 and 6 monolayers (MLs), where 1 ML=0.65 nm. For a quantitative analysis, the sample set is complemented by two reference samples F|Pt, in which NbSe₂ is replaced by Pt (3 nm) providing a large $\theta_{Pt} \approx 10\%$ [21,45]. The in-plane magnetization of all F layers is controlled by an external magnetic field with a strength of about ±10 mT.

*2.2 Methodology*

Our methodology is based on the detection of spin currents $j_s$ and S2C by measuring the emitted THz electric field $E(t)$ (Figure 1a). The F|N samples with F=Ni or Fe and N=NbSe₂ or Pt are excited by near-infrared femtosecond laser pulses (duration of 10 fs, center wavelength of 800 nm, pulse energy of ~1 nJ, repetition rate of 80 MHz) from the substrate side. For clarity and without loss of validity, the schematic (Figure 1a) shows the pump pulse on the opposite side (see above).

As observed previously [24-26,36,38,40,45-53], excitation of F|N stacks leads to an out-of-plane spin current $j_s$ with polarization parallel to the F magnetization **M**, which is converted into an in-plane charge current $j_c$ by S2C, thus, generating an electromagnetic pulse with transient electric field $E(t)$ directly behind the sample. Note that, here, $j_s$ and $j_c$ have the same dimension of m⁻² s⁻¹. To probe $E(t)$, the emitted THz pulse is focused on a ZnTe(110) crystal (thickness of 1 mm) and detected by electro-optic sampling [54,55]. All experiments are performed at room temperature under ambient conditions.

In the frequency domain and in the thin-film approximation [56-58], the complex-valued THz field amplitude is given by

$$E(\omega) = eZ(\omega) \int \mathrm{d}z\, j_c(z,\omega) = eZ(\omega) \int \mathrm{d}z\, \theta(z) j_s(z,\omega). \tag{1}$$

Here, $z$ is the coordinate along the sample normal (Figure 1a), $\omega/2\pi$ is frequency, $Z$ is the total sample impedance, and $\theta(z)$ is the local spin Hall angle that characterizes the strength of S2C. In Equation (1), the total integrated charge sheet current $I_c = \int \mathrm{d}z\, j_c(z)$ has, in principle, contributions from all layers and their interfaces, depending on the local value of $j_s(z)$ and $\theta(z)$. Guided by Figure 1d and Ref. [40], we assume that $\theta(z)$ can be characterized by three S2C values: $\theta_F$ for the F bulk, $\theta_N$ for the N bulk and $\theta_I$ for the F/N interface. It follows that $I_c$ equals the sum

$$I_c = \theta_N J_N + \theta_I J_I + \theta_F J_F, \tag{2}$$

where $J_i = \int_i \mathrm{d}z\, j_s(z)$ is the sheet spin current integrated over the respective layer $i = $ N, I or F. We note that, in this geometry, similar to spin-pumping experiments at GHz frequencies [30], the valley Hall effect is not operative due to the in-plane spin polarization of $j_s$ [18].

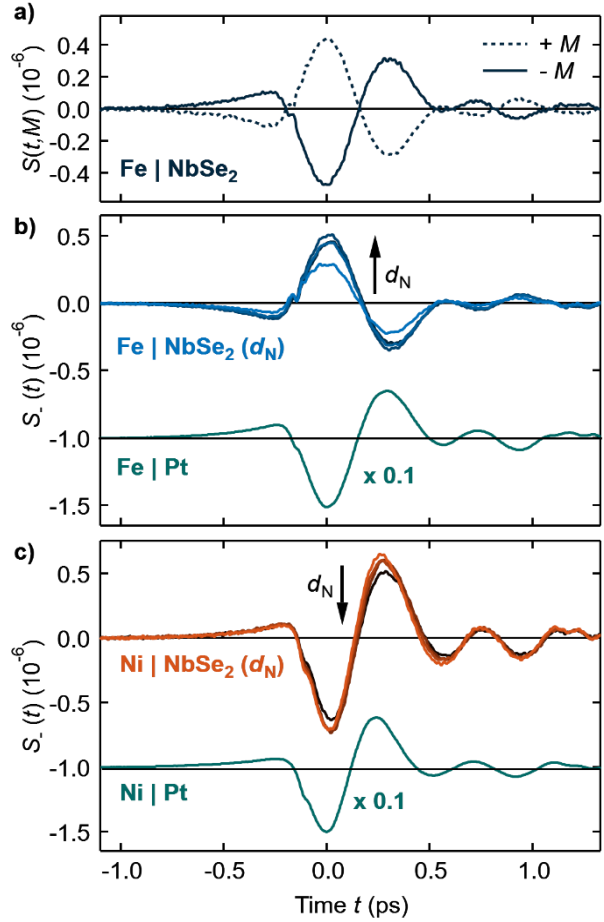

**Figure 2.** THz emission from F|NbSe$_2$ stacks and F|Pt reference samples. a) THz electro-optic signals $S(t, \pm \mathbf{M})$ from Fe|NbSe$_2$(6ML) for opposite orientation of the in-plane Fe magnetization $\mathbf{M}$. b) Antisymmetric component $S_-(t)$ with respect to $\mathbf{M}$ for Fe|NbSe$_2$($d_N$) (upper blue curves) with $d_N = 2, 3, 4, 6$ ML, where the arrow indicates the increasing $d_N$. The signal from the reference sample Fe|Pt (lower green curve) is also shown and vertically offset and scaled by 0.1 for clarity. c) Same as panel b) but for F=Ni.

*2.3 Raw THz emission signals*

Figure 2 shows raw THz waveforms emitted from F|NbSe$_2$ bilayers with varying thickness and for F=Fe or Ni, complemented by a comparison to their respective reference samples F|Pt. A typical THz waveform obtained from Fe|NbSe$_2$(6 ML) is displayed in Figure 2a. Here, the THz signal $S(t, \mathbf{M})$ almost fully reverses upon reversal of the magnetization $\mathbf{M}$ of the Fe layer (solid and dashed waveforms), indicating the magnetic origin of the THz emission. As we are interested in effects odd in $\mathbf{M}$, we focus on the signal $S_-(t) = [S(t, +\mathbf{M}) - S(t, -\mathbf{M})]/2$ in the following. We also observe an even component $S_+(t) = [S(t, +\mathbf{M}) + S(t, -\mathbf{M})]/2$, which is typically an order of magnitude smaller than the odd component (see Figure S1, Supporting Information).

The THz amplitude scales linearly with the energy of the optical pump (Figure S2, Supporting Information). This behavior is typical for a photocurrent in the small-perturbation regime. Reversal of the sample (pump from metallic side as depicted in Figure 1a) leads to reversal of the THz signal (see Figure S3, Supporting Information), consistent with a spin current flowing from F to N [40]. All these observations are confirming the emission scenario summarized in Figure 1a.

Figure 2b-c shows the signals $S_-(t)$ from Fe|NbSe$_2$ and Ni|NbSe$_2$ for $d_\mathrm{N} = 2, 3, 4, 6$ ML and their respective F|Pt references. We make four observations: (i) The THz signals from F|NbSe$_2$ are reversed when F=Fe is replaced by Ni. In contrast, the signals from the reference F|Pt do not reverse. (ii) The amplitude of the emitted waveforms depends on $d_\mathrm{N}$. The trend of the absolute value of the amplitude is, however, opposite for sample sets with F=Fe (Figure 2b) and Ni (Figure 2c), where, respectively, $|S_-|$ increases and decreases with $d_\mathrm{N}$. (iii) The THz signal amplitudes of F|NbSe$_2$ are generally one order of magnitude smaller than those from their F|Pt references. (iv) Apart from this scaling factor, the temporal shape of all traces (Figure 2b-c) is almost identical.

Observation (iv) indicates that the driving force of the THz spin current in F|NbSe$_2$ and F|Pt is the same, that is, a transient spin voltage [35]. Therefore, the root-mean square (RMS) of the THz signal is a good measure of the emission strength. From features (i) and (iii) and by using Eq. (2), we can conclude without any quantitative analysis that S2C in our F|NbSe$_2$ samples is not, unlike in F|Pt, dominated by the non-ferromagnetic layer and, thus, significantly affected by S2C inside F or at the interface. However, observation (ii) suggests a sizable contribution to S2C from the TMDC as well.

*2.4 Normalized signals vs NbSe$_2$ thickness*

To address the impact of the TMDC thickness, we normalize the RMS of the traces in Figure 2b-c by the independently measured impedance $Z$ and pump absorbance $A$ [30,40] for each sample (see Figure S4, Supporting Information). Owing to Equation (1), the resulting quantity is the RMS of the THz sheet charge current $I_\mathrm{c} = \int \mathrm{d}z\, j_\mathrm{c}$ (see Figure 1) normalized by the absorbed pump energy.

The RMS of $I_\mathrm{c}$ is shown for the Fe and Ni sample sets in Figure 3 and reveals a non-trivial trend as a function of $d_\mathrm{N}$. While the dependence $I_\mathrm{c}(d_\mathrm{N})$ is non-monotonic in case of Fe|NbSe$_2$ (Figure 3a), it is monotonically decreasing for Ni|NbSe$_2$ (Figure 3b). The total current $I_\mathrm{c}$ in Fe|NbSe$_2$ is a factor of roughly 1.5 larger than in Ni|NbSe$_2$. If the charge current $I_\mathrm{c}$ was fully generated in F, the normalized $I_\mathrm{c}$ would decrease monotonically with increasing $d_\mathrm{N}$ because less pump-pulse energy would be deposited in F. While such a monotonic decrease is observed for Ni|NbSe$_2$ (Figure 3b), it is not for Fe|NbSe$_2$ (Figure 3a). Thus, at least for Fe|NbSe$_2$, we cannot conclude that $I_\mathrm{c}$ exclusively flows in the ferromagnet or at the F/N interface. In other words, the spin current is injected into the bulk of NbSe$_2$.

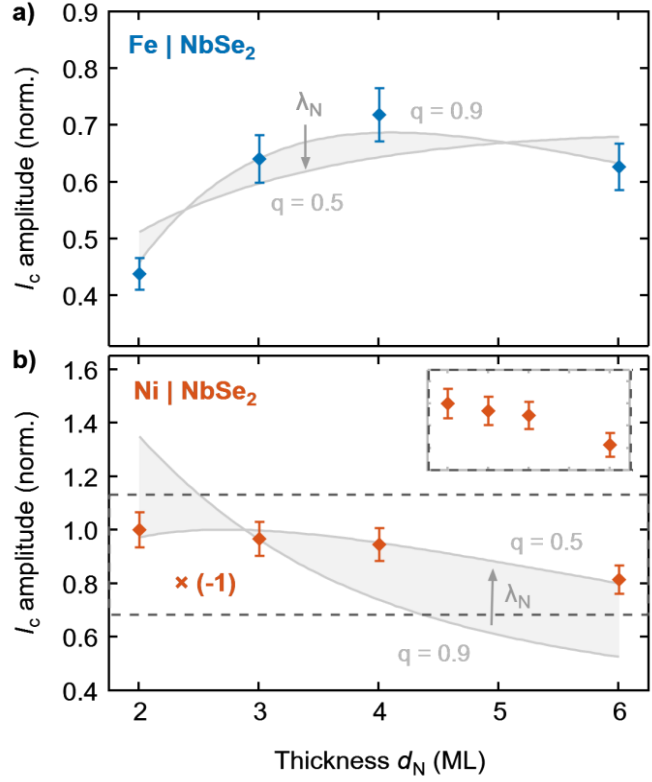

**Figure 3.** Impact of the TMDC layer on total S2C. a) RMS amplitudes of the sheet charge current $I_c$, normalized to the absorbed pump-pulse energy, versus NbSe$_2$ thickness $d_N$ for Fe|NbSe$_2$ and b) Ni|NbSe$_2$ stacks. For clarity, the amplitudes in panel b) are multiplied by −1 and rescaled by a global factor to achieve $I_c(d_N = 2\ \mathrm{ML}) = 1$. The gray areas represent fits according to the transport model for varying spin-current relaxation lengths $\lambda_N = 0.9\text{-}5.8$ nm (indicated by gray arrows) and weighting factor $q = 0.9$. The curve corresponding to the largest $\lambda_N$ also overlaps with the solution for $q = 0.5$ and $\lambda_N = 5.8$ nm (see Table S1).

*2.5 Spin-current model*

To gain further insight into each individual S2C contribution, we model $j_s(z)$, express the resulting $I_c$ using Equation (2) and compare it to the data shown in Figure 3. For this purpose, we assume that the pump pulse drives a spin current $j_s$ that is initially incident on the F/N interface with an amplitude $j_{s0}$ (see Figure 1c and Figure 4a). Subsequently, $j_s$ is partially reflected at the semitransparent F/N interface (spin transmission coefficient $0 < t_s < 1$, reflection coefficient $r_s = t_s - 1$, thus $-1 < r_s < 0$) and decays exponentially with the relaxation length $\lambda_i$. For the NbSe$_2$/substrate interface, we assume total reflection with $r'_s = -1$. From previous works [24,59], we infer $\lambda_F \sim 1$ nm, which is smaller than the F-layer thickness $d_F = 3$ nm. Therefore, we can neglect reflections off the left F boundary (Figure 4a).

By summing up all reflection echoes, we obtain the integrated spin currents $J_N = \int_N dz\, j_s(z)$ and $J_F = \int_F dz\, j_s(z)$ (see Method Section) in the form

$$J_N(d_N) = j_{s0}\lambda_N \frac{\left(e^{d_N/\lambda_N} - 1\right)^2}{e^{2d_N/\lambda_N} - r_s},$$
$$J_F(d_N) = j_{s0}\lambda_F \left[1 - t_s \frac{1}{e^{2d_N/\lambda_N} - r_s}\right]. \quad (3)$$

Here, $j_{s0}$ is the total spin current incident on the F/NbSe$_2$ interface and directly proportional to the pump-induced spin voltage of F. Because the spin voltage, in turn, is proportional to the density of the absorbed pump power in F, we use the scaling $j_{s0} \propto A/(d_F + Bd_N)$, where $B = \mathrm{Im}(n_N^2)/\mathrm{Im}(n_F^2)$ stands for the relative absorptivity of the N and F material [35]. According to previous work, $B$ of our samples is approximately 1 with, however, substantial variation with NbSe2 thickness and quality [60-62]. We now make use of Eq. (2) to investigate the relative importance of the parameters $t_s$, $\lambda_N$, $B$ and $\theta_i$ for the $d_N$-dependence of the normalized charge sheet current density $I_c$ (Figure 3). This approach will eventually provide us with upper and lower limits to the spin Hall angle and $\lambda_N$ of NbSe2.

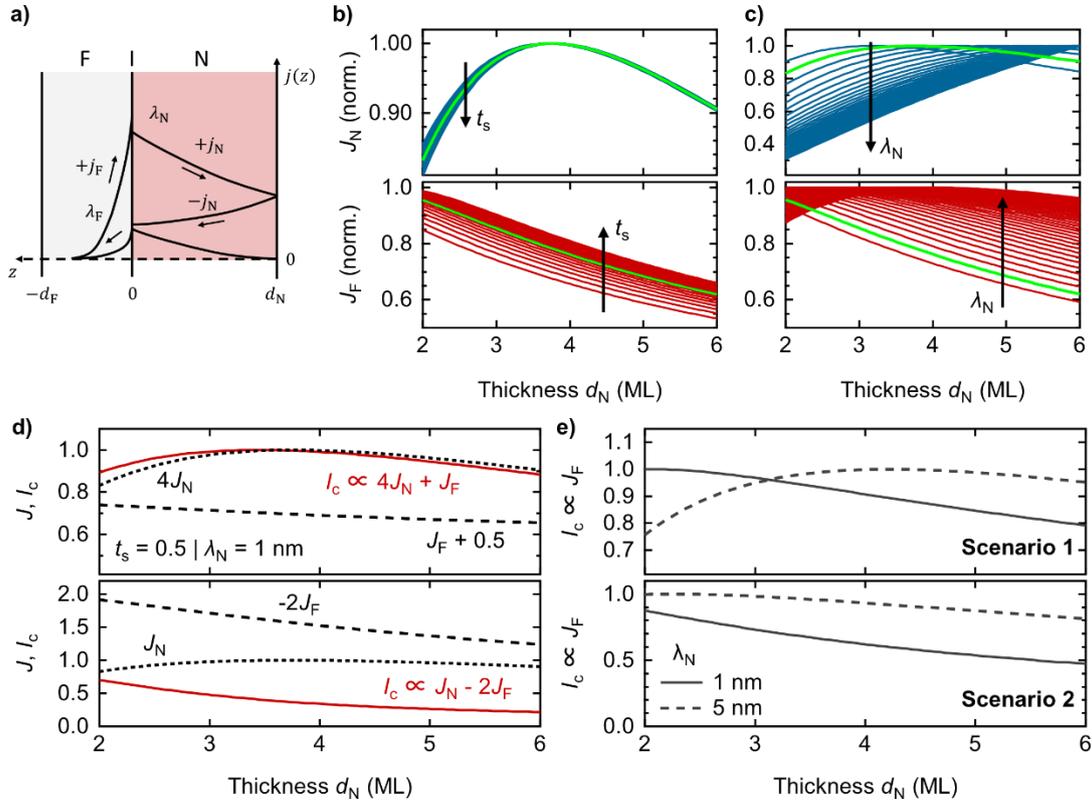

**Figure 4.** Model of spin-current propagation and qualitative analysis. a) Schematic of the $z$-dependence of the spin-current densities $j_N(z)$ and $j_F(z)$, capturing the exponential decay of $j_N(z)$ on the relaxation length $\lambda_N$, and multiple reflections from a semitransparent F/N interface with spin transmission coefficient $t_s$. b) Layer-integrated spin currents in the non-magnetic layer ($J_N$, top panel) and ferromagnet ($J_F$, bottom panel) as a function of the NbSe2 thickness $d_N$ for $\lambda_N = 1$ nm and varying $t_s = 0.1$-$0.9$ (increase depicted by arrows). c) $J_N$ and $J_F$ for $t_s = 0.5$ and varying $\lambda_N = 0.8$-$8$ nm. Curves calculated for default values $t_s = 0.5$, $\lambda_N = 1$ nm and the ratio $B = 1$ of relative absorptivity are plotted in green in panels b) and c). d) Two examples of linear combinations of $J_N$ and $J_F$, both calculated for default values of $t_s$ and $\lambda_N$, leading to different profiles of the total charge current $I_c$. e) Reconstruction of a non-monotonic trend of $I_c$ (scenario 1) and a monotonically decreasing trend (scenario 2) assuming that the transport is totally dominated by the current in F ($I_c \propto J_F$). Calculations are done for $\lambda_N = 1$ nm, where a non-monotonic trend cannot be achieved within the parameter range of $t_s = 0.1$-$0.9$ and $B = 0.5$-$2$ (scenario 1: $t_s = 0.9$, $B = 0.5$; scenario 2: $t_s = 0.5$, $B = 2$). For $\lambda_N = 5$ nm, a monotonic trend is possible (scenario 1: $t_s = 0.8$, $B = 1$; scenario 2: $t_s = 0.4$, $B = 1$).

Figures 4b and 4c show calculated values of $J_N$ and $J_F$ vs the experimental range of $d_N$ for $B = 1$ and various $t_s = 0.1$-$0.9$ and $\lambda_N = 0.8$-$8$ nm, respectively. We see that, for all considered $t_s$, $J_N$ is non-monotonic, whereas $J_F$ keeps its monotonic behavior (Figure 4b). Our calculations indicate a qualitatively similar and, thus, small impact of the variation of $B$. Similarly, spin memory loss at the interface (captured by $a = t_s - 1 - r_s \neq 0$) is a minor effect, too (see Figure S5, Supporting Information). In contrast, variation of $\lambda_N$ has a stronger impact on $J_N$ and $J_F$. For $\lambda_N > 2$ nm, $J_N(d_N)$ changes to a monotonically increasing trend, and a non-monotonic dependence of $J_F(d_N)$ appears for $\lambda_N > 4$ nm in the relevant $d_N$ interval.

Finally, according to Equation (2), the total sheet charge current $I_c$ is determined by a linear combination of the sheet spin current densities $J_N(d_N)$ and $J_F(d_N)$, where $\theta_N$ and $\theta_F$ play the role of weight factors,

$$I_c = J_F(d_N)\theta_F(1 + v_{F/N}) + J_N(d_N)\theta_N. \tag{4}$$

Here, the first term proportional to $J_F$ captures the S2C in both the F bulk and at the F/NbSe$_2$ interface, where $v_{F/N} = d_I/\lambda_F$ denotes the relative contribution of the interface with effective thickness $d_I$. To stress this fact, we can define an effective spin Hall angle $\theta_F' = \theta_F(1 + v_{F/N})$ that includes the bulk and interfacial S2C. An illustration of qualitatively different results with non-monotonic and monotonically decreasing trends vs $d_N$ for different linear combinations $I_c = 4J_N + J_F$ and $I_c = J_N - 2J_F$ and default values $t_s = 0.5$ and $B = 1$ is shown by the black solid curves in Figure 4d. This graph is the basis for further qualitative interpretation of our measurements.

*2.6 Qualitative comparison*

We first limit our discussion to cases with $\lambda_N \leq 4$ nm. Such an assumption is realistic because larger THz spin-current relaxation lengths in spintronic THz emission were not observed so far, including materials with negligible spin-orbit interaction like Cu [24,28,40,45,59]. In this case, $J_F$ shows only a monotonically decreasing trend with $d_N$ (Figure 4b-c) and cannot explain the measured non-monotonic $d_N$-dependence of Fe|NbSe$_2$ (Figure 3a). Thus, $I_c$ requires a relatively strong contribution from $J_N$, implying that the spin Hall angle $\theta_N$ is comparable or larger than $\theta_F'$. In other words, there is a sizable component of S2C in the bulk of NbSe$_2$.

To confirm the robustness of this qualitative conclusion, we assume that both Fe- and Ni-based sample sets may have different model parameters $t_s$ and $B$. By varying these parameters within realistic intervals 0.1-0.9 and 0.5-2, respectively, we aim to obtain a non-monotonic trend of $I_c$ vs $d_N$, as observed in Fe-based samples, without the contribution of $J_N$ (taking $\theta_N = 0$). As documented by the solid lines in Figure 4e (scenario 1 and 2), the results cannot capture the required dependence without considering a sizable $\theta_N$, even for the extreme values of $t_s$ and $B$.

The situation may change if we assume $\lambda_N > 4$ nm. In this case, $J_F$ vs $d_N$ can reach a non-monotonic trend. By choosing sufficiently different model parameters for scenario 1 ($t_s = 0.8$, $B = 1$) and scenario 2 ($t_s = 0.4$, $B = 1$), we can model qualitatively the same $d_N$-dependence of $I_c$ without any contribution of $J_N$. A similar discussion with respect to variations of the spin-memory-loss parameter leads to a qualitatively identical conclusion as for the variation of $t_s$ (see Figure S5, Supporting Information).

To summarize, our modeling can reproduce the measured $d_N$-dependence of the normalized $I_C$ (Figure 3) for only two scenarios: In scenario 1, the spin current is converted in the bulk of the TMDC with sizable $\theta_N$ and has a relaxation length $\lambda_N < 4$ nm. In scenario 2, there is no significant S2C in the TMDC, the spin current has $\lambda_N > 4$ nm, and the spin transmission coefficients of the Fe/NbSe$_2$ and Ni/NbSe$_2$ interfaces differ by a factor of 2.

*2.7 Fitting of $I_c(d_N)$*

After identifying two theoretically possible scenarios of spin current propagation, we can now roughly estimate the absolute magnitude of $\theta_N$ and $\lambda_N$ by simultaneously fitting Equation (4) to both data sets of the $A$-normalized $I_c(d_N)$ in Figures 3a and 3b. The fit minimizes the weighted sum of the squares $R^2$ of both data sets $qR^2_{Fe|NbSe_2} + (1-q)R^2_{Ni|NbSe_2}$ with the relative weight $q$ that allows us to prioritize one of the trends. Because the normalized data are determined up to a global scaling factor, the fitting process yields only the relative proportion $\theta_{NbSe_2} : \theta'_{Fe} : \theta'_{Ni}$. To obtain the absolute value of the spin Hall angles, we compare it to the $A$-normalized $I_c$ of the reference samples Fe|Pt and Ni|Pt (Figure 2b-c) and use $\theta_{Pt} \approx 10\%$ [21], as detailed in the Method Section.

To minimize the number of fitting parameters and simplify the model, we fix the default values of $t_s = 0.5$, $B = 1$, $a = 0$ for both the Fe- and Ni-based sample sets. We first evaluate the fits for selected values of $\lambda_N = 0.9$-5.8 nm. They are shown in Figure 3 by the light-gray-shaded area, whose boundaries represent the extremal $\lambda_N$. To better capture the characteristic non-monotonic trend of the Fe-based sample set, we chose $q = 0.9$. A more detailed description of the fitting results, including more values of $\lambda_N$ and $q$, is given in Table S1 and Figure S6 (Supporting Information). Consistent with our qualitative observations above, the inferred spin Hall angle $\theta_{NbSe_2}$ ranges from about $-1.1\%$ (for $\lambda_N = 0.9$ nm) down to approximately $-0.2\%$ ($\lambda_N > 4$ nm). If we allow $\lambda_N$ to be a free parameter, the fits yield solutions depending on $q$, as shown in Table 1 and by the dark grey boundary curves in Figure 3 for $q = 0.5$ and

0.9. They support our qualitative model insights: The transport is either characterized by $\theta_{\mathrm{NbSe_2}} \approx -1\%$ and $\lambda_{\mathrm{N}} \sim 1$ nm, or a negligible $\theta_{\mathrm{NbSe_2}}$ and a rather unusually large $\lambda_{\mathrm{N}} \approx 6$ nm [24,28,40,45,59].

**Table 1.** Results of the fitting procedure of the $A$-normalized $I_c(d_N)$ in Figures 3a and 3b. The extracted spin Hall angles are $\theta_{\mathrm{NbSe_2}}$ for bulk NbSe$_2$, $\theta'_F$ for bulk F=Fe or Ni including the conversion at the F/NbSe$_2$ interface, for different choices of $q$. The values of the spin Hall angles in the bottom row are stable within the single-digit decimal precision over the entire interval of $q$. The uncertainty of the values is roughly 10%.

|  | $\theta_{\mathrm{NbSe_2}}$ (%) | $\theta'_{\mathrm{Fe}}$ (%) | $\theta'_{\mathrm{Ni}}$ (%) | $\lambda_{\mathrm{N}}$ (nm) |
| --- | --- | --- | --- | --- |
| $q = 1.0$ | $-1.5$ | $0.5$ | – | $0.8$ |
| $q = 0.9$ | $-1.1$ | $0.3$ | $1.5$ | $0.9$ |
| $0.1 < q < 0.8$ | $-0.2$ | $-0.5$ | $1.2$ | $5.5$-$6.1$ |

2.8 Discussion

The inferred values of $\theta'_{\mathrm{Ni}}$ and $\theta'_{\mathrm{Fe}}$ are in reasonable agreement with literature magnitudes of around 1% [21,40,63]. While our analysis confirms the expected $\theta'_{\mathrm{Ni}} > 0$ [24,44,45,59], the small value of $\theta'_{\mathrm{Fe}}$ does not allow for a clear statement on the sign of $\theta'_{\mathrm{Fe}}$, considering that these quantities contain also the interfacial contribution of possibly similar magnitude but possibly different polarity [40].

To further discuss the inferred spin Hall angles of NbSe$_2$, we performed *ab-initio* calculations of the spin Hall conductivity $\sigma_{\mathrm{SH}}$ for an out-of-plane-propagating spin current with in-plane spin polarization (see details in the Method Section and calculations in Figure S7 of the Supporting information). We obtain $\sigma_{\mathrm{SH}} = -20$ S/cm at the Fermi level and a local minimum of $-31$ S/cm at 0.13 eV below. Considering the mean measured conductivity of $\sigma_{\mathrm{NbSe_2}} = 2.2 \times 10^3$ S/cm in Fe|NbSe$_2$ and Ni|NbSe$_2$ (see Figure S4, Supporting Information), the expected spin Hall angle $\theta_{\mathrm{NbSe_2}} = \sigma_{\mathrm{SH}}/\sigma_{\mathrm{NbSe_2}}$ lies in the interval $-(0.9$-$1.4)\%$. This value, including its sign, is excellently consistent with the value $\theta_{\mathrm{NbSe_2}} \approx -1\%$ extracted from our experiment (Figure 3) for scenario 1 with the shorter $\lambda_{\mathrm{N}} \approx 1$ nm.

Even though our experiment and theory agree well, we emphasize that our quantitative analysis has to be taken with caution because of a potentially large uncertainty of model parameters and non-trivial model assumptions. Formulated more conservatively, we conclude that the spin Hall angle of NbSe$_2$ is negative and has a magnitude larger than 1%. The corresponding THz spin-current relaxation length $\lambda_{\mathrm{N}}$ is of the order of a few nanometers.

We note that the extracted values might also be affected by a possible contribution of magnetic-dipole radiation to the THz signal, which arises from ultrafast demagnetization [35]. However, the $d_{\mathrm{N}}$-dependence and opposite polarity of the emitted signal in Fe and Ni-based sample sets (Figure 2) rule out a dominant role of this effect in the total signal. This notion is corroborated by the fact that well reversed THz signals are found when the samples are rotated by 180° about the magnetization (see Figure S3, Supporting Information). They indicate that a maximum contribution of magnetic-dipole radiation of 20% to the THz signal is possible. By rescaling and fitting the respective amplitude data in Figure 3, the quantitative analysis yields the same values as summarized in Table 1 within the given error bars of 10%. Therefore, we consider the possible impact of the magnetic-dipole radiation negligible for our experiment.

**3. Conclusions**

We performed broadband THz emission spectroscopy on ultra-high-vacuum-grown epitaxial stacks composed of NbSe$_2$ with varying thickness and a ferromagnetic metal. Using a qualitative analysis of the emitted THz pulses after optical excitation, we infer that the in-plane spin-polarized current is injected from the ferromagnet into the TMDC on ultrafast time scales. By comparison of the different thickness dependence of the emission from Fe- vs Ni-based sample sets to a spin-current model, we conclude that there is either a spin-to-charge current conversion in the bulk of TMDC with THz spin current relaxation length of ∼ 1 nm, or no significant conversion happens but the THz spin current relaxation length in the TMDC has a less realistic value above 4 nm. A quantitative analysis, based on fitting the model to the measured data, confirms the qualitative notion and yields the spin Hall angle of NbSe$_2$ in the range −(1.1-0.2)% with corresponding spin current relaxation lengths of 1-6 nm. Our findings show that ultrafast spin-current injection into the TMDC NbSe$_2$ is possible and that broadband THz emission spectroscopy [27,36,38,41,45,64] is an excellent and versatile tool for such investigations that complements the established THz methods [56,65,66].

**4. Method Section**

*Sample preparation:* Atomically thin films of NbSe$_2$ (2, 3, 4, 6 ML) were deposited by the hybrid-PLD [42,43]. Double-polished *c*-cut sapphire substrates were used as a substrate. The growth temperature was 500°C as monitored by an infrared pyrometer working around a wavelength of 10 μm. Fluxes of pure Nb (99.9%) and Se (99.999%) were applied for a duration calculated from a growth rate of ~10 min/ML established by measuring the thickness of a reference sample using X-ray reflectivity (XRR) at a synchrotron light source [42]. The films were annealed at 400°C under Se flux for 1 h after growth. Subsequently, the sample was capped with Se (15 nm) at room temperature. The capped samples were transferred through ambient air into another chamber, where the Se capping layer was removed by heating (150°C). Subsequently, magnetic films of Fe or Ni (3 nm) were deposited by electron-beam evaporation. Finally, the heterostructure films were capped *in-situ* with Al (2 nm) which readily oxidized in air to serve as an AlOx protection layer.

*Spin-current model:* To model spin-current propagation in a F|N stack with a semitransparent F/N interface, we consider the configuration in Figure 1. The F|N layers are stacked along the coordinate $z$ with origin set at the interface: the TMDC (layer N, thickness $d_\text{N}$) is located at $z > 0$, the ferromagnet (layer F, thickness $d_\text{F}$) at $z < 0$. The interface is characterized by the spin-current transmission coefficient $0 < t_\text{s} < 1$ and reflectivity coefficient $r_\text{s} = -1 + t_\text{s} < 0$. The reflection at the right sample boundary (N side) is considered using the coefficient $r'_\text{s} < 0$. As the spin-current relaxation length fulfills $\lambda_\text{F} \ll d_\text{F}$ in the F layer, we do not account for any effect of the left boundary.

By considering first a very thick N layer ($\lambda_\text{N} \ll d_\text{N}$), the spin current $j_{\text{s}0}$ created at $z = 0$ after the optical excitation follows an exponential profile $j_\text{s}(z) = j_{\text{s}0} e^{z/\lambda_\text{F}}$ in F ($z < 0$) and $j_\text{s}(z) = j_{\text{s}0} e^{-z/\lambda_\text{N}}$ in N ($z > 0$).

If we consider realistic thicknesses of N and allow for back-reflections off the N boundary at $z = d_\text{N}$, the echos forms an infinite sequence, yielding in N, i.e., for $z \in [0, d_\text{N}]$,

$$\frac{j_\text{s}(z)}{j_{\text{s}0}} = e^{-z/\lambda_\text{N}} + r'_\text{s} e^{-2d_\text{N}/\lambda_\text{N}} e^{z/\lambda_\text{N}} + r'_\text{s} r_\text{s} e^{-2d_\text{N}/\lambda_\text{N}} e^{-z/\lambda_\text{N}} + (r'_\text{s} r_\text{s})^2 e^{-4d_\text{N}/\lambda_\text{N}} e^{z/\lambda_\text{N}} + \cdots =$$
$$= \frac{1}{1 - q_\text{N}} \left[ \exp\left(-\frac{z}{\lambda_\text{N}}\right) + r'_\text{s} \exp\left(-\frac{2d_\text{N} - z}{\lambda_\text{N}}\right) \right] \tag{S1}$$

where

$$q_\text{N} = r'_\text{s} r_\text{s} e^{-2d_\text{N}/\lambda_\text{N}}.$$

Similarly, in F, i.e., for $z \in [-d_\text{F}, 0]$, we have

$$\frac{j_\text{s}(z)}{j_{\text{s}0}} = \left[ 1 + \frac{t_\text{s} r'_\text{s}}{1 - q_\text{N}} \exp\left(-\frac{2d_\text{N}}{\lambda_\text{N}}\right) \right] \exp \frac{z}{\lambda_\text{N}}. \tag{S2}$$

This solution still fulfils continuity of $j_\text{s}(z)$ at $z = 0$. The layer-integrated spin currents in N, F and at the interface (I) are obtained by

$$J_\text{N}(d_\text{N}) = \int_0^{d_\text{N}} dz\, j_\text{s}(z) = j_{\text{s}0} \frac{\lambda_\text{N}}{1 - q_\text{N}} \left[ 1 - \exp\left(-\frac{d_\text{N}}{\lambda_\text{N}}\right) \right] \left[ 1 + r'_\text{s} \exp\left(-\frac{d_\text{N}}{\lambda_\text{N}}\right) \right], \tag{S3a}$$

$$J_\text{F}(d_\text{N}) = \int_{-\infty}^0 dz\, j_\text{F}(z) = j_{\text{s}0} \lambda_\text{F} \left[ 1 + \frac{t_\text{s} r'_\text{s}}{1 - q_\text{N}} \exp\left(-\frac{2d_\text{N}}{\lambda_\text{N}}\right) \right], \tag{S3b}$$

$$J_\text{I}(d_\text{N}) = \int_{-d_\text{I}}^{0} dz\, j_\text{F}(0) = J_\text{F}(d_\text{N}) d_\text{I}/\lambda_\text{F}. \tag{S3c}$$

In the last integral, we assume that the interface is described by a very thin interlayer of effective thickness $d_\text{I}$ over which the spin current $j_\text{I}$ is approximately constant and equals $j_\text{I} = j_s(0)$.

By taking $r_s' = -1$, we obtain Equation (3). A possible spin memory loss $a$ is included in the model by considering that $t_s$ and $r_s$ do not add up to unity: $t_s - r_s = 1 + a$ where $-1 < a < 0$, and by taking $-1 < r_s' < 0$.

*Determination of spin Hall angles:* The proportion $\theta_\text{NbSe2}:\theta_\text{Fe}':\theta_\text{Ni}'$ of the spin Hall angles, but not their absolute value, can be determined by fitting Equation (4) with the corresponding $\theta_\text{F}'$ and calculated $J_\text{F}$, $J_\text{N}$ on two sets of data in Figure 3. To estimate the value of the spin Hall angles, the data were complemented by simultaneous measurements of two reference samples Fe|Pt and Ni|Pt. Similarly, to data in Figure 3, the RMS of their signals $S(t)$ were normalized to the corresponding pump light absorptance $A = 59.8\,\%$ and $61.5\,\%$ and impedance $Z = 40.5\,\Omega$ and $38.9\,\Omega$, respectively, obtaining normalized $I_c^\text{Fe|Pt}$ and $I_c^\text{Ni|Pt}$.

In accordance to Equation (1), we can expect that the S2C in these reference samples will be dominated by conversion in Pt ($\theta_\text{F} \ll \theta_\text{Pt} \approx 10\,\%$) and, thus, [24]

$$I_c^\text{F|Pt} = e\theta_\text{Pt} J_\text{Pt} = e\theta_\text{Pt}\lambda_\text{Pt} \frac{1}{d_\text{F} + d_\text{Pt}} \tanh\frac{d_\text{Pt}}{2\lambda_\text{Pt}}, \tag{S4}$$

where $J_\text{Pt}$ is the integrated spin current in Pt, including a possible back-reflection using $r_s = r_s' = -1$, and other quantities with the same meaning as in the main text. For both sets with F=Fe and Ni and N=NbSe₂, we obtain the following two equations with two unknowns $\theta_\text{F}'$ and $\theta_\text{N}$:

$$\frac{I_c^\text{F|N}}{I_c^\text{F|Pt}} = C_\text{F} = \frac{\theta_\text{F}' J_\text{F} + \theta_\text{N} J_\text{N}}{\theta_\text{Pt} J_\text{Pt}}, \tag{S5a}$$

$$D_\text{F} = \frac{\theta_\text{N}}{\theta_\text{F}'}. \tag{S5b}$$

Here, $C_\text{F}$ are experimental inputs, $D_\text{F}$ are known from fitting, spin currents $J_\text{F}$, $J_\text{N}$ and $J_\text{Pt}$ are calculated using Equations (S3) and (S4), $\theta_\text{Pt} \approx 10\,\%$ and $\lambda_\text{Pt} = 1$ nm are taken as reference values from literature [21,40,45]. Since we use only the ratio of simultaneously measured $I_c$ in Equation (S5a), all F-specific factors affecting the emission process are canceled out. Solving Equation (S5) for both F yields

$$\theta_\text{N} = C_\text{F}\theta_\text{Pt} \frac{J_\text{Pt}}{J_\text{F}/D_\text{F} + J_\text{N}}, \tag{S6a}$$

$$\theta_\text{F}' = C_\text{F}\theta_\text{Pt} \frac{J_\text{Pt}}{J_\text{F} + J_\text{N}/D_\text{F}}. \tag{S6b}$$

*Ab-initio calculations:* Density functional calculations are performed for the bulk NbSe₂ with an in-plane lattice constant of 3.44 Å. The distance between the van der Waals layers is 2.89 Å. The generalized gradient approximation of Perdew-Burke-Ernzerhof [67] is used for the exchange correlation potential as implemented in the FLEUR code [68]. The maximally localized Wannier functions are constructed using the WANNIER90 code in conjunction with the FLEUR package [69,70], based on which an effective Hamiltonian in a tight-binding scheme is constructed for the calculation of the DC spin Hall conductivity $\sigma_{ij}^l$ according to

$$\sigma_{ij}^l = e\hbar \int \frac{d^3k}{(2\pi)^3} \sum_{n=1}^{N_\text{occ}} \Omega_{n,ijl}^S(\mathbf{k}), \tag{S7}$$

$$\Omega_{n,ijl}^S(\mathbf{k}) = -2\text{Im} \sum_{m \neq n} \frac{\langle\psi_{m\mathbf{k}}|J_i^{l,s}|\psi_{n\mathbf{k}}\rangle\langle\psi_{n\mathbf{k}}|v_j|\psi_{m\mathbf{k}}\rangle}{(E_{n\mathbf{k}} - E_{m\mathbf{k}})^2}. \tag{S8}$$

Here, $\Omega_n^S(\mathbf{k})$ is the spin Berry curvature of all occupied states. $v_j$ is the $j$-th Cartesian component of the velocity operator, $|\psi_{n\mathbf{k}}\rangle$ is the Bloch function of band $n$ at wavevector $\mathbf{k}$ with energy $E_{n\mathbf{k}}$, and $J_i^{l,s} = (\hbar/2)\{\sigma_l, v_i\}$ describes a spin current flowing into i direction with spin polarization along the $l$-th axis.

**Supporting Information**

Supporting Information is available from the Wiley Online Library or from the author.

**Acknowledgements**

The authors acknowledge funding by the German Research Foundation through the collaborative research centers SFB TRR 227 "Ultrafast spin dynamics" (Project ID 328545488, projects A05, B02 and B10), the European Union H2020 program through the projects CoG TERAMAG (Grant No. 681917) and ASPIN (Grant No. 766566), and the Czech Science Foundation through Project GA CR (Grant No. 21-28876J). H.N. thanks I. Lakemeyer for her support in preparing the magnetic films.

**Conflict of Interest**

The authors declare no conflict of interest.

**Data Availability Statement**

The data that support the findings of this study are available from the corresponding author upon reasonable request.

## Supporting information

## Terahertz spin-to-charge current conversion in stacks of ferromagnets and the transition-metal dichalcogenide NbSe₂

Lukáš Nádvorník, Oliver Gueckstock, Lukas Braun, Chengwang Niu, Joachim Gräfe, Gunther Richter, Gisela Schütz, Hidenori Takagi, Tom S. Seifert, Peter Kubaščík, Avanindra K. Pandeya, Abdelmadjid. Anane, Heejun Yang, Amilcar Bedoya-Pinto, Stuart S. P. Parkin, Martin Wolf, Yuriy Mokrousov, Hiroyuki Nakamura, Tobias Kampfrath

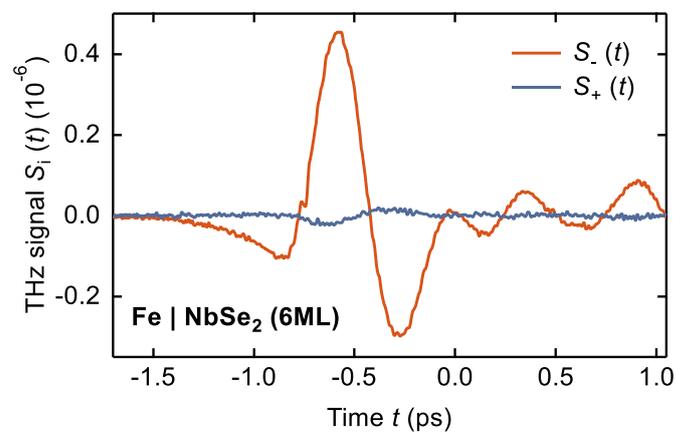

**Figure S1.** Comparison of magnetic and non-magnetic contributions to the THz emission signal. THz waveforms of the magnetic contribution $S_-(t) = [S(t, +\mathbf{M}) - S(t, -\mathbf{M})]/2$ (orange) and the non-magnetic component $S_+(t) = [S(t, +\mathbf{M}) + S(t, -\mathbf{M})]/2$ (violet) are shown for Fe|NbSe₂(6 ML). All non-magnetic signal contributions are considerably smaller than the magnetic ones.

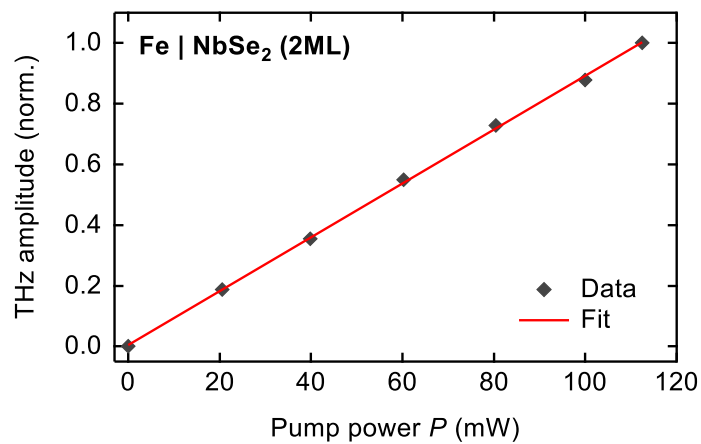

**Figure S2.** Fluence dependence of the THz emission signal for Fe|NbSe₂. THz emission amplitude of Fe|NbSe₂(2 ML) as a function of pump power (gray diamonds), normalized to the maximum value. The red line is a linear fit with intercept zero.

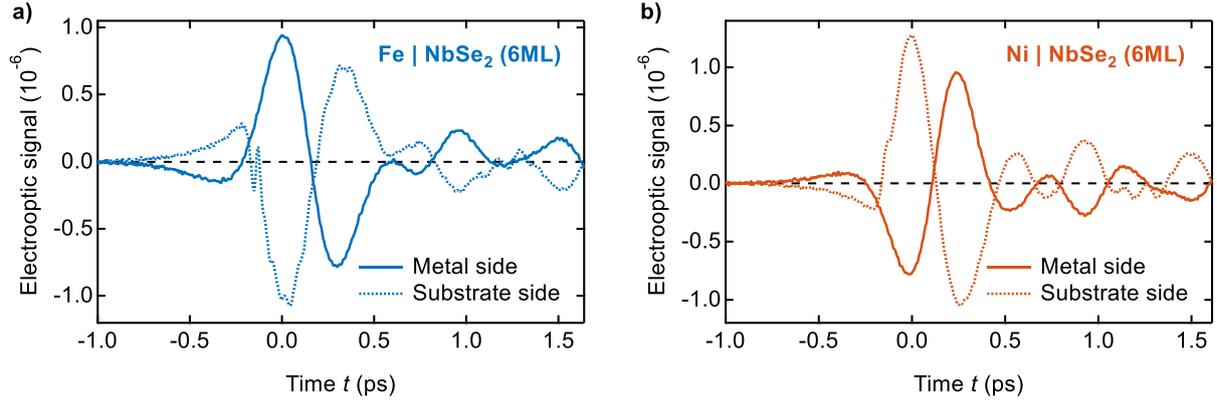

**Figure S3.** Impact of sample turning. Normalized THz emission signals of a) Fe|NbSe$_2$(6 ML) and b) Ni|NbSe$_2$(6 ML) for optical excitation from the substrate (dashed line) and the metal side (solid line). The possible impact of magnetic-dipole radiation due to ultrafast demagnetization of Fe and Ni can be estimated by symmetrizing the respective pairs of curves, yielding < 5% and < 20% in panels a and b, respectively.

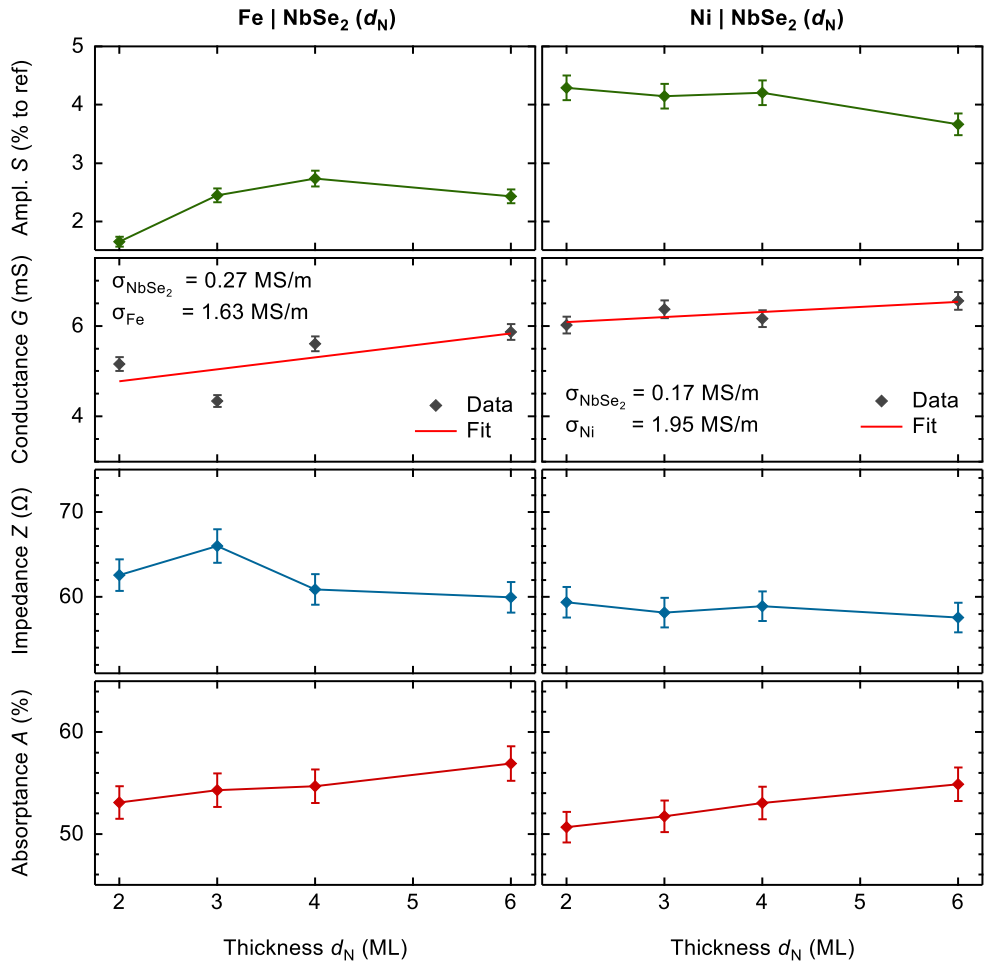

**Figure S4.** Optical and electrical properties of F|NbSe$_2$($d_N$) samples. Left panels: F=Fe, amplitude of the THz signal trace relative to the F|Pt reference sample, the sheet conductance $G$, the real part of the impedance $Z = Z_0/(Z_0 G + n_1 + n_2)$, where $Z_0 = 377\ \Omega$, and absorbtance $A$ of the pump beam (center wavelength at 800 nm) as functions of thickness $d_N$ of the NbSe$_2$ layer. Right panels: same as on the left-hand side, but for F=Ni. $G$, and consequently $Z$, were obtained from electrical measurements using the Van der Pauw method.

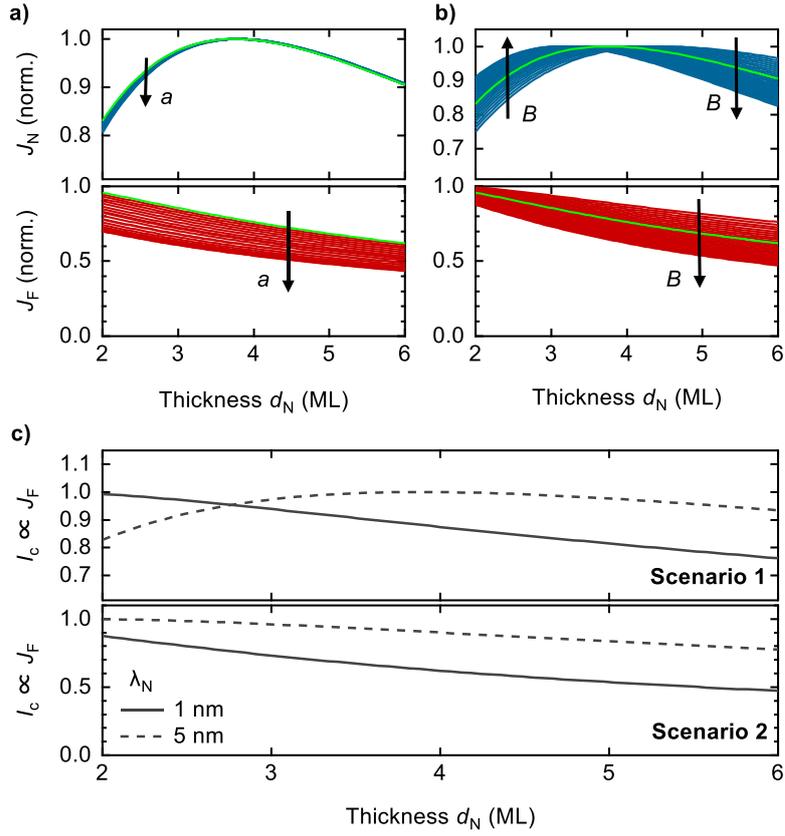

**Figure S5.** Model of spatial spin-current distribution. a) Spin sheet currents, i.e., layer-integrated spin-current densities, in the non-magnetic layer ($J_N$, top panel) and in the ferromagnet ($J_F$, bottom panel) as a function of the N thickness $d_N$ for $t_s = 0.5$ and spin memory loss $a$ varying between 0 and 0.9, where the increase is indicated by arrows. b) $J_N$ and $J_F$ for $j_0 = 1/(1 + B d_N/d_F)$ and $B$ varying from 0.5 to 2. Curves calculated for default values $a = 0$ and $B = 1$ are plotted in green in panels a and b, respectively. c) Reconstruction of a non-monotonic trend of the total sheet charge current $I_c$ (scenario 1) and monotonically decreasing trend (scenario 2) without any S2C in NbSe$_2$ for $\lambda_N = 1$ nm or 5 nm, but with default values of $a$ and $t_s$ as specified above.

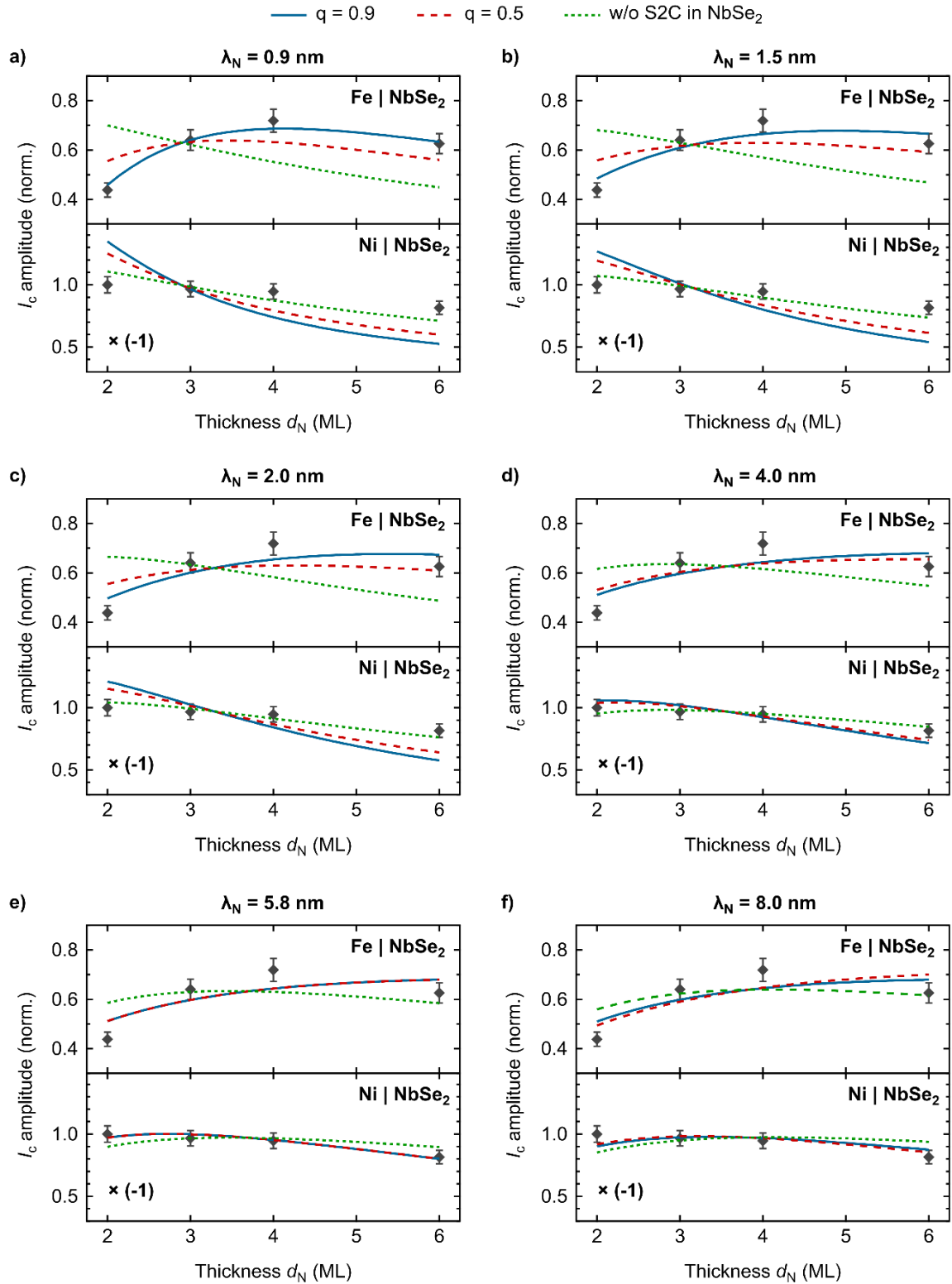

**Figure S6.** Fitting of data from Figure 3 (grey scatters) by Equation (4) for $q = 0.9$ (blue, solid curve) and 0.5 (red, dashed line) for various fixed values of $\lambda_N = $ a) 0.9 nm, b) 1.5 nm, c) 2.0 nm, d) 4.0 nm, e) 5.8 nm, f) 8.0 nm. Green dotted curves represent fits by Equation (4) when assuming no S2C in the bulk of NbSe$_2$ ($\theta_N = 0$).

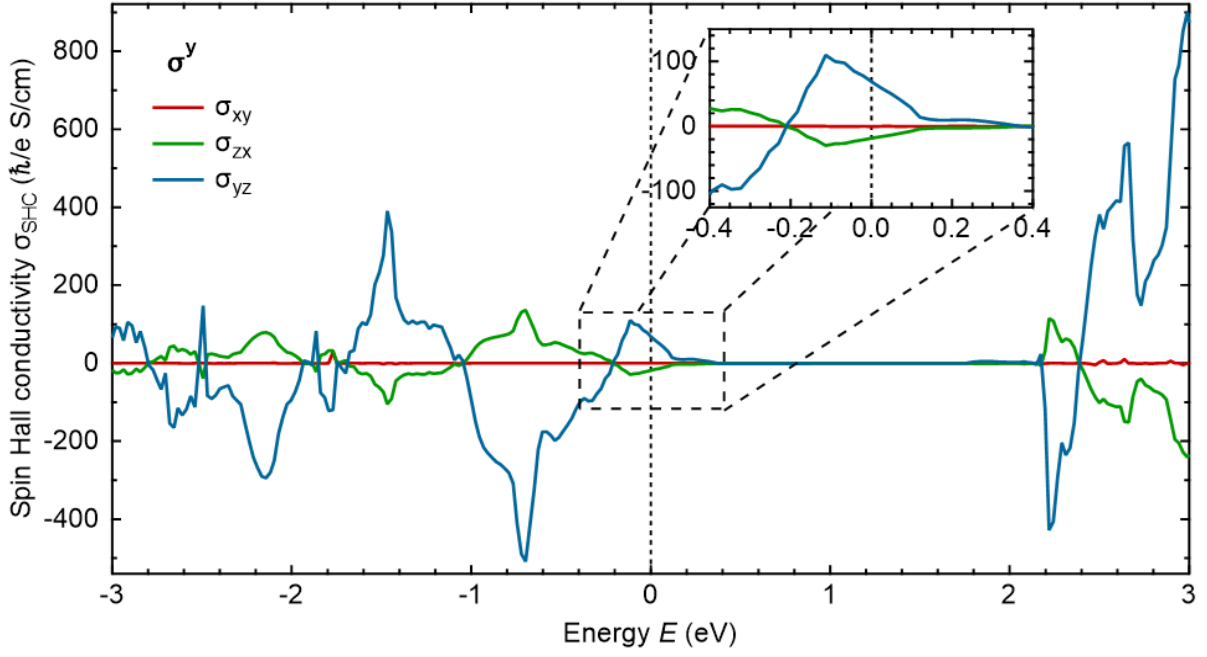

**Figure S7.** *Ab-initio* calculation of the spin Hall conductivity of NbSe$_2$. Calculated DC spin Hall conductivity $\sigma_{ij}^l$ as function of single-electron energy $E$, where $i$ refers to the spin current direction, $j$ to the direction of the applied electric field, and $l$ denotes the polarization of the spin current. In our study, we measure spin currents in the $z$ direction ($i = z$) with an emitted THz electric field in $x$ direction ($j = x$) and an in-plane spin polarization ($l = y$). Therefore, the relevant spin Hall conductivity is shown by $\sigma_{zx}^y$ (green curve). The inset shows a magnified detail around the Fermi energy at $E = 0$, indicated with a dashed horizontal line.

| $\lambda_N$ (nm) | $q = 0.9$ | | | | $q = 0.5$ | | | |
|---|---|---|---|---|---|---|---|---|
| | $\theta_{NbSe2}$ (%) | $\theta'_{Fe}$ (%) | $\theta'_{Ni}$ (%) | Error ($10^{-3}$) | $\theta_{NbSe2}$ (%) | $\theta'_{Fe}$ (%) | $\theta'_{Ni}$ (%) | Error ($10^{-3}$) |
| 0.8 | −1.34 | 0.43 | 1.59 | 5.25 | −0.81 | 0.08 | 1.26 | 18.50 |
| 0.9* | −1.09 | 0.31 | 1.47 | 5.19 | −0.66 | 0.00 | 1.19 | 17.92 |
| 1.0 | −0.88 | 0.20 | 1.37 | 5.26 | −0.54 | −0.06 | 1.13 | 17.17 |
| 1.5 | −0.49 | −0.03 | 1.18 | 6.09 | −0.31 | −0.20 | 1.02 | 13.67 |
| 2.0 | −0.36 | −0.14 | 1.12 | 6.58 | −0.24 | −0.27 | 1.01 | 10.64 |
| 3.0 | −0.26 | −0.27 | 1.12 | 6.62 | −0.19 | −0.34 | 1.04 | 6.60 |
| 4.0 | −0.21 | −0.36 | 1.15 | 6.39 | −0.17 | −0.40 | 1.11 | 4.57 |
| 5.0 | −0.18 | −0.46 | 1.20 | 6.21 | −0.16 | −0.46 | 1.18 | 3.74 |
| 5.8** | −0.16 | −0.49 | 1.24 | 6.13 | −0.16 | −0.50 | 1.24 | 3.58 |
| 6.0 | −0.16 | −0.51 | 1.25 | 6.12 | −0.16 | −0.51 | 1.26 | 3.59 |
| 7.0 | −0.14 | −0.58 | 1.31 | 6.11 | −0.16 | −0.56 | 1.34 | 3.84 |
| 8.0 | −0.12 | −0.65 | 1.37 | 6.15 | −0.16 | −0.61 | 1.42 | 4.33 |

**Table S1.** Fit parameters for various values of $\lambda_N$ and two weight factors $q$. Rows indicated by * and ** are fit optima for $q = 0.9$ and $q = 0.5$, respectively. The uncertainties of the fit parameters follow from the presented sensitivity analysis in the table.